\documentclass[11pt,twoside]{article}

\usepackage{asp2006}
\usepackage{epsf}
\usepackage{lscape}

\begin{document}
\title{Dynamical Influence of Bars on the Star Formation in Isolated Galaxies}   %%% Fill in title
\author{S. Verley$^{1,2,3}$, F. Combes$^{1}$, L. Verdes-Montenegro$^{2}$, G. Bergond$^{2,4}$, S. Leon$^{5,2}$}   %%% Fill in author names
\affil{$^{1}$LERMA, $^{2}$IAA, $^{3}$INAF, $^{4}$GEPI, $^{5}$IRAM}    %%% Fill in author affiliations

\begin{abstract} %%% Abstract to run on from here.
   Star formation depends strongly on both the local environment of galaxies and the internal dynamics of the interstellar medium. To disentangle the two effects, we obtained, in the framework of the AMIGA project, H$_\alpha$ and Gunn r photometric data for more than 200 spiral galaxies lying in very low-density regions of the local Universe.
   We characterise the H$_\alpha$ emission, tracing current star formation, of the 45 largest and least inclined galaxies observed for which we estimate the torques between the gas and the bulk of the optical matter. We subsequently study the H$_\alpha$ morphological aspect of these isolated spiral galaxies.
   Using Fourier analysis, we focus on the modes of the spiral arms and also on
the strength of the bars, computing the torques between the gas and newly formed stars (H$_\alpha$), and the bulk of the optical matter (Gunn r).
\end{abstract}

%%% MAIN BODY OF TEXT GOES HERE. CONSULT "INSTRUCTIONS FOR AUTHORS USING
%%% LATEX2E MARKUP", SECTIONS 2.3-2.6 FOR HELP WITH EQUATIONS, FIGURES,
%%% AND TABLES.

%\section{}   %%% Top level section head (remove "%" symbol)
%\subsection{}   %%% Second level section head (remove "%" symbol)
%\subsubsection{}   %%% Lowest level section head (remove "%" symbol)
%\section*{}    %%% Unnumbered top level section head (remove "%" symbol)
%\subsection*{}   %%% Unnumbered second level section head (remove "%" symbol)

\section{The H$_\alpha$ AMIGA sample of isolated spiral galaxies}
One objective of our project was to study the different aspects of H$_\alpha$ emission from isolated spiral galaxies. With this aim, and in order to avoid well
known biases due to flux- or magnitude-limited samples and to end with a complete and homogeneous sample, we kept all the galaxies in a volume-limited sample, i.e. with observed recession heliocentric velocities $V_{\rm R}$:

\begin{center}
1500 km\,s$^{-1}$ $\le$ $V_{\rm R}$ $\le$ 5000 km\,s$^{-1}$.
\end{center}

\noindent For the study presented here (45 galaxies) two further requirements were imposed: major axis $a \ge 1'$ to have a sufficient spatial resolution; inclination $i \le 50^\circ$ in order to obtain a sufficiently accurate deprojection.

\section{Bars in isolated galaxies}

The barred galaxies represent 60\% of our sample while the unbarred galaxies represent 33\% of the sample, and galaxies with intermediate stage bar represent 7\% of the total: isolated galaxies span the whole range of bar morphologies, in quantities similar to the galaxies in denser environments. Isolated galaxies are not preferentially barred or unbarred. We also noticed the existence of a shift angle between the location of the gas and the location of the older stellar component: the H$_\alpha$ emission was always leading with respect to the bar in the Gunn r image; the orientations of the H$_\alpha$ and stellar bars correspond to shift angles of typically 10$^\circ$.

\section{Secular evolution experienced by isolated galaxies}

We interpreted the various global H$_\alpha$ morphologies observed in terms of the secular evolution experienced by galaxies in isolation. The main H$_\alpha$ classes can be related to the bar evolution phases. The observed frequency of particular patterns brings constraints on the time spent in the various evolution phases. We defined three main groups: group {\bf E} (19 galaxies) consists of the galaxies showing a strong central peak in the H$_\alpha$ emission, H$_\alpha$ emission along the spiral arms but not in the bar; group {\bf F} (9 galaxies) comprises galaxies with less gas, which do not present any central emission knot in H$_\alpha$; galaxies in group {\bf G} (8 galaxies) show H$_\alpha$ emission in the bar.

Other groups with fewer galaxies were also defined, most noticeable is the group {\bf EG} (3 galaxies) featuring galaxies that have characteristics in between those defined by the groups {\bf E} and {\bf G}. Hence, we could interpret these features as different stages of an evolutive sequence: {\bf G} $\to$ {\bf EG} $\to$ {\bf E} $\to$ {\bf F}.\\

Numerical simulations showed a predicted frequency of the {\bf G} phase higher than the {\bf E} phase, in contradiction with our observations. We attribute this discrepancy to a failure of the star formation recipe, since we used the usual local Schmidt law for the star formation rate. The frequently observed phenomenon of star formation  avoiding the bar, in spite of large gas density there, suggests that the star formation law should also depend on other factors, in particular the relative velocity of the gas in the bar.

\section{Conclusions}
We interpret the various bar/spiral morphologies observed in terms of the secular evolution experienced by galaxies in isolation. We also classify the different spatial distributions of star forming regions in barred galaxies. The observed frequency of particular patterns brings constraints on the lifetime of the various evolution phases. We propose an evolutive sequence accounting for the transitions between the different phases we observed. Isolated galaxies do not appear to be preferentially barred or unbarred. Fitting the H$_\alpha$ distributions using numerical simulations yields constraints on the star formation law, which is likely to differ from a genuine Schmidt law. In particular, it is probable that the relative velocity of the gas in the bar also needs to be taken into account.

\acknowledgements %%% Text of acknowledgements runs on after this command.
We wish to thank the organizing committees for this interesting conference.

%%% THE BIBLIOGRAPHY
%%%
%%% CONSULT SECTION 3 OF "INSTRUCTIONS FOR AUTHORS" FOR HOW TO USE NATBIB.
%%% AUTHORS ARE ENCOURAGED TO USE EITHER THE "THEBIBLIOGRAPY" ENVIRONMENT
%%% BY UNCOMMENTING (DELETING THE "%" SYMBOL) THE COMMANDS BELOW, OR BY
%%% USING THE BIBTEX ENVIRONMENT. TO FIND OUT WHICH IS APPLICABLE TO YOUR
%%% CONTRIBUTION, CONSULT THE VOLUME EDITORS FOR YOUR PROCEEDINGS.
%%%

\end{document}